\documentclass[aps,prx,twocolumn,secnumarabic]{revtex4-2} 

\usepackage[utf8]{inputenc}
\usepackage{longtable}
\usepackage{morefloats}
\usepackage[dvips]{graphicx,color}
\usepackage[dvipsnames]{xcolor}
\usepackage{epsfig,amsfonts,amsbsy}
\usepackage{amsmath,amsthm,amssymb}
\usepackage{bbold}
\usepackage{url}
\usepackage{verbatim}
\usepackage{array}
\usepackage{multirow}
\usepackage{tabularx}
\usepackage{braket} 
\usepackage{dsfont}
\usepackage{bm}
\usepackage{natbib}
\usepackage{multibib}
\usepackage{physics}
\usepackage[colorlinks=true, pdfstartview=FitV, linkcolor=blue,
citecolor=blue, urlcolor=blue]{hyperref}
\usepackage[capitalise]{cleveref}

\usepackage{textcomp}

\begin{document}

\title{Creating entangled logical qubits in the heavy-hex lattice with topological codes}
\author{Bence Het\'enyi}
\email{bence.hetenyi@ibm.com}
\author{James R. Wootton}
\affiliation{IBM Quantum, IBM Research Zurich, Switzerland} 
\date{\today}

\begin{abstract}
Designs for quantum error correction depend strongly on the connectivity of the qubits. For solid state qubits, the most straightforward approach is to have connectivity constrained to a planar graph. Practical considerations may also further restrict the connectivity, resulting in a relatively sparse graph such as the heavy-hex architecture of current IBM Quantum devices. In such cases it is hard to use all qubits to their full potential. Instead, in order to emulate the denser connectivity required to implement well-known quantum error correcting codes, many qubits remain effectively unused. In this work we show how this bug can be turned into a feature. By using the unused qubits of one code to execute another, two codes can be implemented on top of each other, allowing easy application of fault-tolerant entangling gates and measurements. We demonstrate this by realizing a surface code and a Bacon-Shor code on a 133 qubit IBM Quantum device. Using transversal CX gates and lattice surgery we demonstrate entanglement between these logical qubits with code distance up to $d = 4$ and five rounds of stabilizer measurement cycles. The nonplanar coupling between the qubits allows us to simultaneously measure the logical $XX$, $YY$, and $ZZ$ observables. With this we verify the violation of Bell’s inequality for both the $d=2$ case with post selection featuring a fidelity of $94\%$, and the $d=3$ instance using only quantum error correction.
\end{abstract}

\maketitle

\section{Introduction}
\begin{figure*}
    \centering
    \includegraphics[width = \textwidth]{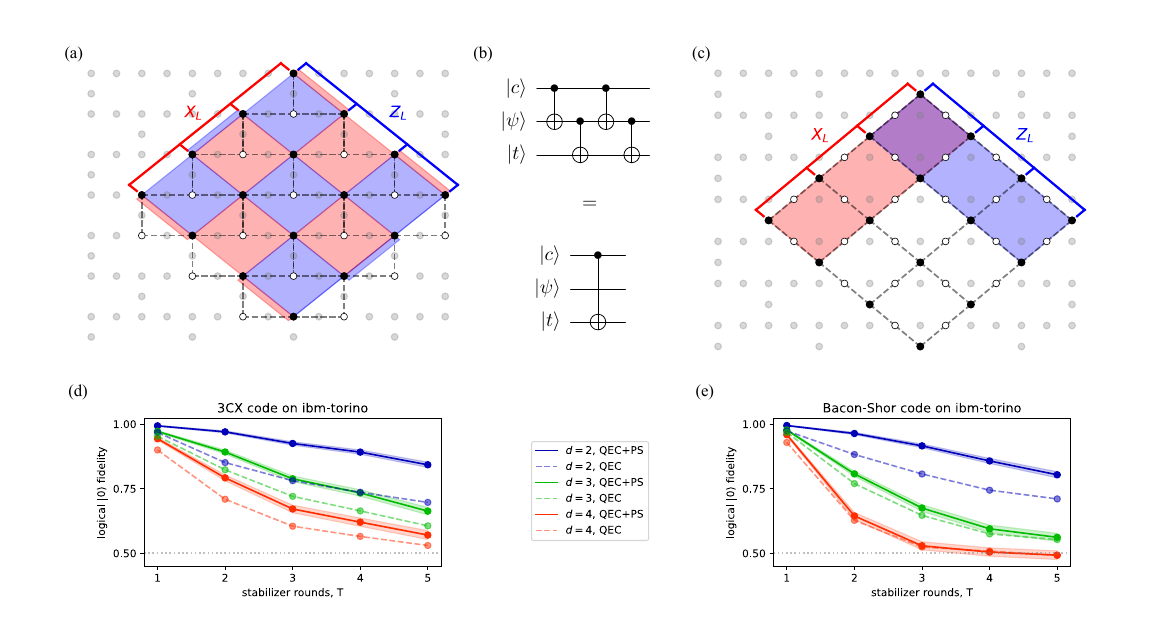}
    \caption{(a) and (c) qubit sublattices used for the 3CX and the Bacon-Shor codes at $d=4$, respectively. Black (white) dots are code (ancilla) qubits, red and blue tiles represent the $X$ and $Z$ stabilizers (only two out of six are shown for the BS code), and dashed lines indicate the effective next-nearest neighbour connectivity required for the stabilizer measurements. (b) next-nearest neighbour CX gate implemented via four CX gates without affecting the middle qubit $\ket{\psi}$. (d) and (e) logical error rate as a function of stabilizer rounds for multiple code distances decoded using minimum-weight perfect matching with (QEC+PS) and without (QEC) post-selecting the ambiguous matchings.}
    \label{fig:HHXlattice}
\end{figure*}

Fault-tolerant quantum computing requires several ingredients. Firstly, a scalable hardware with high-quality physical qubits that can carry the burden of the large qubit overhead imposed by quantum error correcting (QEC) codes. Error rates of the  physical qubits need to be below a threshold error rate, such that logical errors can be arbitrarily suppressed with increasing code distance $d$. Superconducting qubit platforms are approaching this regime as indicated by quantum memory experiments on single logical qubits (LQs)~\cite{krinner2022realizing,google2023suppressing}. On the other hand, in order to perform fault-tolerant quantum computation in the future, one also needs to demonstrate single-qubit control and entanglement between logical qubits.

A logical Bell state with a $d=2$ surface code has been performed with trapped ions using lattice surgery (with fidelity of $\mathcal F = 75\%$ using post-selection) \cite{erhard2021entangling} and even higher fidelities have been achieved with the $d=3$
color code using transversal CX and error correction \cite{postler2022demonstration, dasilva2024demonstration}. Recently neutral atoms have been the first platform where entangled logical qubits were prepared for multiple code distances with the surface code (with an estimated fidelity of $93\%$ for $d=7$)~\cite{bluvstein2023logical}. On the other hand, experimental realization of entangled logical qubits with superconducting quantum processors have been hitherto limited to low-distance non-topological codes \cite{harper2019fault,gupta2024encoding}. Furthermore, there is no demonstration of logical qubit entanglement to date over various code distances and stabilizer measurement rounds in any qubit platform.

Logical gates typically have large qubit overhead for topological error correcting codes. In planar architectures, as currently seen for superconducting qubits, transversal CX gates are usually not feasible, as the SWAP overhead would rapidly grow with the code distance. Braiding of hole- or twist defects require additional space in the qubit lattice in order to maintain the code distance during the operation~\cite{brown2017poking}. Even lattice surgery, which has a modest qubit overhead, would require $\sim 150$ physical qubits to implement a logical CX gate between two distance-5 logical qubits in the surface code~\cite{horsman2012surface,vuillot2019code}.

Quantum error correction on the heavy-hexagonal lattice faces the additional challenge of the low connectivity. Flag qubit-based QEC codes proposed for this connectivity typically offer lower error thresholds than the surface code~\cite{chamberland2020topological}. A recently proposed variant of the surface code, know as the 3CX surface code, can be implemented on a hexagonal connectivity without a major compromise in performance~\cite{mcewen2023relaxing}. However, the qubits at the links of the heavy-hex lattice remain unused or utilized as flag qubits \cite{mcewen2023relaxing,benito2024comparative}.

In this work we realize the 3CX surface code on a 133 qubit IBM Quantum device, and show that link qubits can be used to simultaneously implement a Bacon-Shor code~\cite{shor1995scheme,bacon2006operator,egan2021fault} on the same patch of physical qubits. Since the stabilizer operators of the 3CX code are compatible with those of the Bacon-Shor code, one can utilize transversal CX gates as well as lattice surgery to establish and probe the entanglement between the two logical qubits~\cite{poulsen2017fault,gidney2023bacon}. We execute such schemes for multiple code distances ($d \in \{2,3,4\}$) and several rounds (up to $T=5$), demonstrating logical qubit entanglement for the first time in a superconducting device using topological QEC codes. 

The nonplanar connectivity between the two codes allows for simultaneous $XX$ and $ZZ$ lattice surgery measurements, indirectly measuring the $YY$ logical, which is sufficient to calculate the fidelity and verify the violation of Bell's inequality. We report violation of Bell's inequality using both error correction and post selection, with fidelities up to $\mathcal F = 93.7\%$ in the latter case. 

The paper is organized as follows. In Sec.~\ref{sec:singleLQ} we show how the 3CX and the Bacon-Shor codes can be implemented on the heavy-hex lattice and present the data on the logical performance of each code. Sec.~\ref{sec:twoLQgates} concerns itself with entangling the two logical qubits and probing the resulting logical Bell-state. We confirm the fault-tolerance of our protocol and the scaling of logical error rate through simulations in Sec.~\ref{sec:thresholds}. Finally, we present the experimental data on the entangled logical qubits in Sec.~\ref{sec:exp_fidelity}. The experimental and simulated results are supported by additional details in Apps.~\ref{app:more_thresholds}-\ref{app:detector_circuits}.

\section{3CX and Bacon-Shor codes on the heavy-hex lattice}
\label{sec:singleLQ}

In this work we directed our attention to two topological codes, the 3CX surface code \cite{mcewen2023relaxing} and the Bacon-Shor code \cite{bacon2006operator}. While none of these codes fit directly with the heavy-hexagonal connectivity of IBM devices \cite{chamberland2020topological}, by making a few observations we can easily embed both of them to this coupling map simultaneously with a constant qubit overhead. Dividing the heavy-hex lattice into two sublattices, one with vertices only, and another including only the links, we get a hexagonal lattice and a heavy-square grid --i.e., having an extra qubit on every edge of a square grid-- as shown in  Fig.~\ref{fig:HHXlattice}(a) and (c). Effective next-nearest neighbour CX gates can be implemented using four CXs of the original connectivity regardless of the state of the middle qubit (see Fig.~\ref{fig:HHXlattice}(b)). This scheme provides means to couple the qubits of each sublattice, without affecting the qubits of the other sublattice.

The 3CX code inherits the high error threshold and well-established tools to implement logical gates from the rotated surface code~\cite{mcewen2023relaxing,hetenyi2023tailoring}. Even accounting for the connectivity of the heavy-hex lattice, the threshold error rate is $\sim 0.3\%$ both with and without flag qubits~\cite{benito2024comparative}. Our method with the next-nearest-neighbour CX gates matches this performance, as shown in App.~\ref{app:more_thresholds}, while freeing up the link qubits. This is a major improvement compared to the heavy-hex code, which has no threshold in the thermodynamic limit and its pseudo-thresholds are below $0.1\%$.

The 127- and 133-qubit IBM Quantum devices can accommodate the 3CX code up to code distance $d=4$ (see Fig.~\ref{fig:HHXlattice} (a)). See App.~\ref{app:RBdata} for the benchmark data corresponding to the 133-qubit device. Performing stabilizer measurements for up to $T=5$ rounds, we perform error correction using a standard minimum-weight perfect-matching decoder~\cite{higgott2023sparse}. The results are shown in Fig.~\ref{fig:HHXlattice}(d). While the code is clearly operating above threshold, the fidelity of the logical $\ket{0}_L$ state, i.e., $\mathcal F = \bra{0}_L\rho_\text{exp}\ket{0}_L$, remains above $50\%$ even after five rounds for the highest code distance ($d=4$). Post selection of the results for ambiguous matchings further enhances the logical performance reaching $99.3\%$ in the case of $d=2$ and $T=1$.

In order to use the full potential of the even-distance variants, we employ a post-selection strategy, discarding shots where the matching implies error strings of length $d/2$. To this end, we introduce a trivial boundary node on the edge opposite to the logical edge and check if the optimal matchings include the same number of edges when matching to the trivial (logical) boundary is enforced~\cite{hutter2014efficient,gidney2023yoked}. The fact that the post-selection notably affects the logical error rates of the $d=3$ data as well indicates that the $(d+1)/2$-long error strings are not yet the dominant source of logical errors for the noise levels of the device.

On the other hand, the Bacon-Shor code can be naturally implemented on a heavy-square lattice. While this code has no threshold in the thermodynamic limit, logical error rates below $10^{-8}$ can be achieved using lattice surgery (LS) and code concatenation, without modifying the connectivity requirements of the code~\cite{gidney2023bacon}. Here, due to the restricted code distance, we focus on the traditional variant without concatenation~\cite{shor1995scheme,bacon2006operator}. In this case, we do require the bearded edge of the latest 133-qubit device to implement the $d=4$ version of the code (lower edge on Fig.~\ref{fig:HHXlattice}(c)). The corresponding results with and without post-selection are presented in Fig.~\ref{fig:HHXlattice}(e). Although the error rates at low code distance match that of the 3CX code ($99.4\%$ in the case of $d=2$ and $T=1$ with post selection), the decay of fidelity is faster for increasing code distances. Nonetheless, the BS code is still a valuable addition to the set of QEC codes that are compatible with the heavy-hex lattice.

Interleaving the two codes gives rise to correlated errors between the two codes, resulting from CX errors. Decomposing the resulting hyperedges we obtain, slightly reduced threshold values. e.g., $p_\text{th} = 0.2\%$ for the 3CX code, and similarly reduced pseudo-threshold values for the BS code. Importantly, however, we maintain the correct scaling of the logical error rate with physical error rate for both codes (see App.~\ref{app:more_thresholds}). Experimental data for the simultaneous execution of the two codes can be found in App.~\ref{app:additional_data}.

\section{Entangling logical qubits with transversal CXs and lattice surgery}
\label{sec:twoLQgates}

In this section we show that overlaying the 3CX and the BS codes allows us to use both transversal CX gates and the available lattice surgery tools of the BS code~\cite{poulsen2017fault,gidney2023bacon} thereby entangling the logical qubits of the two codes. In the following subsections we first describe the stabilizers of the 3CX and the BS codes and show that their logical subspaces are compatible, i.e., a transversal CX gate (CX$_\perp$) can be used to entangle the two LQs from these two different codes. We identify some special cases, when the codes can be decoded during the CX$_\perp$ with conventional tools using \texttt{qiskit-qec}, \texttt{stim} and \texttt{pymatching}~\cite{gidney2021stim,higgott2023sparse}.

\subsection{Compatibility of the 3CX and the Bacon-Shor codes}

{\it Logical subspace.} Taking a $d\times d$ square grid of code qubits, the logical operators of both the BS and the 3CX codes can be defined as Pauli $X$ ($Z$) operators acting on any row (column) of physical qubits (e.g., the first one was chosen for each logical on Fig.~\ref{fig:HHXlattice}(a) and (c)). Any odd-element product of these equivalent logical operators is also a valid logical operator of the same type. Therefore acting on every physical qubit with the same Pauli, {\it transversally}, is also a logical operation for odd code distance $d$.

{\it Stabilizer generators.} Since the even-element product of any (logical or physical) Pauli is the identity, even products of valid logical  operators are logical identities, i.e., stabilizer operators. In fact, the stabilizer group of the BS code is defined as the product of pairs of Pauli-$X$($Z$) logicals acting on adjacent rows (columns) of qubits (as shown in Fig.~\ref{fig:HHXlattice}(c)). While these stabilizers are also included in the stabilizer group of the 3CX code, the group of the latter is richer containing $d^2-1$ plaquette operators which act on four nearest-neighbor qubits that are located on the corners of an elementary unit cell. At the boundaries of the 3CX lattice weight-two stabilizers need to be introduced that commute with the bulk stabilizers as well as the logicals (see Fig.~\ref{fig:HHXlattice}(a)). There are two equivalent choices of boundary stabilizers: e.g., the first row can either start with a boundary (as in Fig.~\ref{fig:HHXlattice}(a)) or a full plaquette and the rest of the boundaries are unambiguous. The 3CX code alternates round-by-round between these two options, however, BS-like stabilizers remain valid stabilizers in both subcycles.

{\it Gauge groups.} The stabilizer generators of the BS code are highly nonlocal. In practice one would rather measure a set of local observables that commute with the logicals as well as the stabilizers, even if they do not commute with each other. I.e., the individual measurements have random outcomes, but certain products of them (the stabilizers) are deterministic. Such a group of operators is called the gauge group. As we have seen before, the stabilizers of the 3CX code are gauges in the BS code, however, the lowest-weight gauge group is a set of two-qubit parity measurements. E.g., the $X$ gauges are Pauli-Xs acting on every nearest-neighbour qubit pair ($d(d-1)$) that live on adjacent rows.

We conclude this subsection by noting that the 3CX code can be downgraded to a BS code on-demand by changing the stabilizer measurements that are compared in subsequent rounds. In such a round transversal logical CX can be applied between the two codes. After the logical gate stabilizer measurements can continue according to the respective stabilizer groups.

\subsection{Stabilizers and detectors during the transversal CX gate}
\label{sec:detectors}

Detectors are comparisons of gauge or stabilizer measurements that are deterministic in the absence of errors. For the BS code this means a comparison of the same subsets of gauge measurements round-by-round. In the case of the 3CX code the modified CX schedule of the stabilizer measurements, that reduces the required connectivity, transforms the stabilizer group in a nontrivial way such that $X$ ($Z$) stabilizers of subsequent rounds are shifted by one site in their respective row (column)~\cite{mcewen2023relaxing}. Transversally applied CX gates also transform the joint stabilizer group of the two codes in a nontrivial way. In order to ensure that the detectors are deterministic in the absence of errors we need to compare each stabilizer outcome before the CX$_\perp$ with the image thereof after the operation.

Let us define the downgraded stabilizer group of the 3CX code as $\tilde{\mathcal S}^\text{3CX} = \tilde{\mathcal S}^\text{3CX}_X + \tilde{\mathcal S}^\text{3CX}_Z$ where the $X$ and $Z$ subgroups are
\begin{subequations}
\begin{eqnarray}
    \tilde{\mathcal S}^\text{3CX}_X = \{\prod_j S_{ij}\, |\, S_{ij} \in \mathcal S^\text{3CX}_X \And S_{ij} \in i\text{th row} \}
\end{eqnarray}
\begin{eqnarray}
    \tilde{\mathcal S}^\text{3CX}_Z = \{\prod_j S_{ij}\, |\, S_{ij} \in \mathcal S^\text{3CX}_Z \And S_{ij} \in i\text{th column} \}
\end{eqnarray}
\end{subequations}
with $\mathcal S^\text{3CX}_{X(Z)}$ being the $X$ ($Z$) stabilizer subgroup of the 3CX code. The transversal CXs transform a $Z$ stabilizer $S^\text{BS}_{Z,i}$ of the BS code according to the stabilizer flow, or equivalently
\begin{equation}
    \text{CX}_\perp 
    \left( 
        I^\text{3CX} \otimes S^\text{BS}_{Z,i}  
    \right)
     = 
    \left( 
    \tilde{S}^\text{3CX}_{Z,i}
    \otimes S^\text{BS}_{Z,i}
    \right) 
    \text{CX}_\perp.
\label{eq:CXSZCZ_stabs}
\end{equation}
Therefore the measurement outcome of the stabilizer $S^\text{BS}_{Z,i}$ before the CX$_\perp$ needs to be compared to the product of $S^\text{BS}_{Z,i}$ and $\tilde{S}^\text{BS}_{Z,i}$. Similarly, the BS-like $X$ stabilizers of the 3CX code after the CX$_\perp$ read as
\begin{equation}
    \text{CX}_\perp\!
    \left(
        \tilde{S}^\text{3CX}_{X,i}
        \otimes I^\text{BS}
    \right)  =
    \left(
    \tilde{S}^\text{3CX}_{X,i}
    \otimes S^\text{BS}_{X,i}
    \right)
    \text{CX}_\perp.
    \label{eq:CXSZCX_stabs}
\end{equation}

Finally, $X$ stabilizers on the BS code (i.e., generators of $\mathcal S^\text{BS}_X$) and the original $Z$ stabilizers of the 3CX code (generators of $\mathcal S^\text{3CX}_Z$) remain unchanged. Therefore the $X$ detectors of the BS and the $Z$ detectors of the 3CX code are not affected by $\text{CX}_\perp$ if controlled on the data qubits of the 3CX code.

\begin{figure}
    \centering
    \includegraphics[width=0.49\textwidth]{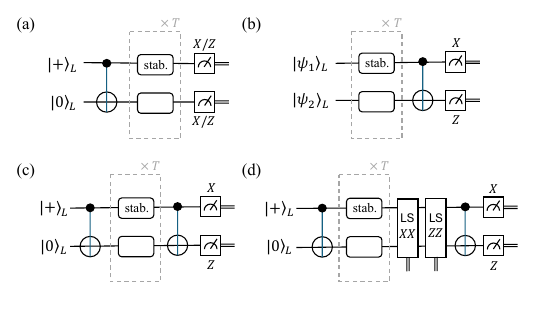}
    \caption{(a) transversal preparation of a Bell state followed by stabilizer measurements and arbitrary state tomography. (b) transversal measurement on the Bell basis, i.e., every 3CX-BS qubit pair is measured in the Bell basis. (c) transversal preparation and measurement on the Bell basis. Logical qubits are disentangled at the point of the measurement. (d) transversal preparation and measurement of a Bell state, with logical-$XX$ and $ZZ$ measurement on the entangled qubits using lattice surgery.}
    \label{fig:Bell_meas_prep_circuits}
\end{figure}

In both the 3CX and the BS code any single $X$ or $Z$ error is detected by a pair of detectors making them amenable for decoding via perfect-matching. In general, the transversal CX gate does not conserve this property, inducing up to four detection events of $Z$ or $X$ type. E.g., let us consider a phase-flip error in the $(i+1)$th row of data qubits in the BS code right after CX$_\perp$. This will be detected by the two conventional $X$ detectors of the BS code, when comparing stabilizers $S^\text{BS}_{X,i}$ (or $S^\text{BS}_{X,i+1}$) between subsequent rounds. Simultaneously, however, it will also be detected by a pair of $X$ detectors where the outcome of $\tilde{S}^\text{3CX}_{X,i}\otimes S^\text{BS}_{X,i}$ after CX$_\perp$ is compared to that of $\tilde{S}^\text{3CX}_{X,i}$ before it (or similarly for $i+1$). In the Tanner-graph representation of the decoding problem, such an error corresponds to a hyperedge connecting the four nodes corresponding to the four detectors that observed it. This issue has also been pointed out in Ref.~\cite{bluvstein2023logical, cain2024correlated} and further supporting information using the detecting region picture of Ref.~\cite{mcewen2023relaxing} is available in App.~\ref{app:hyperedges}. 

There are two special cases, however, for which some stabilizers that contribute to the excess of detection events (i.e., two out of four) are not in the stabilizer group at the time of applying CX$_\perp$. E.g., when the 3CX (control) LQ is prepared in the $\ket{+}_L = \ket{+}^{\otimes d^2}$ state and the BS (target) in the $\ket{0}_L= \ket{0}^{\otimes d^2}$ state before the first stabilizer measurement (see Fig.~\ref{fig:Bell_meas_prep_circuits}(a)) which we call {\it transversal Bell-state preparation}. The stabilizer group of the freshly initialized LQs consists only of the $X$ ($Z$) stabilizers of the 3CX (BS) code. After the $\text{CX}_\perp$ both of these stabilizer types will spread to the other code resulting in a set of delocalized stabilizers as in Eq.~\eqref{eq:CXSZCZ_stabs} and \eqref{eq:CXSZCX_stabs}. These delocalized BS-like stabilizers are the only valid stabilizers in the group, and every qubit touches at most two from each type, resulting in a matching graph without any hyperedges (these arguments are further elaborated on in App.~\ref{app:detectiong_regions}). Fig.~\ref{fig:Bell_meas_prep_circuits}(b) shows the second special case which is --in a stabilizer-flow sense-- the time-reversed partner of the previous, i.e., the {\it transversal Bell-state measurement}. That is when the final measurement in the $X$ ($Z$) basis for the 3CX (BS) code is performed right after $\text{CX}_\perp$. Note that the measurement in an arbitrary basis in Fig.~\ref{fig:Bell_meas_prep_circuits}(a) is not analogous to the first case and it can still induce hyperedges depending on the measurement basis of the given logical as in Ref.~\cite{bluvstein2023logical}.

\subsection{Figures of merit for the logical Bell pair}

Let us define some quantities of interest for the two LQs to identify which logical expectation values are necessary to demonstrate the entanglement. Taking the two-qubit density matrix $\rho_{L12}$, the fidelity of the Bell-state $\ket{\phi_1} = (\ket{00}+\ket{11})/\sqrt 2$ can rewritten as
\begin{eqnarray}
\begin{split}
    \mathcal F_\text{Bell} &= \text{Tr}(\ket{\phi_1} \bra{\phi_1} \rho_{L12}) \\
    & = \frac 1 4 (1+ \ev{X_{L1}X_{L2}} - \ev{Y_{L1}Y_{L2}} + \ev{Z_{L1}Z_{L2}}), 
\end{split}
\end{eqnarray}
highlighting the importance of measuring the logical $\ev{YY}$ expectation value. The second figure of merit is the violation of the CHSH inequality. The formulation based on Bell's inequality suggests that the state should be measured in a $45^{\circ}$ rotated basis, however Ref.~\cite{horodecki1995violating} showed that the left-hand side of the inequality can be evaluated using the matrix of expectation values
\begin{eqnarray}
    (T_\rho)_{ij} = \ev{P^i_1P^j_2}\, ,
\end{eqnarray}
where $P^i_1 \in \{ X_1, Y_1, Z_1\}$ are Pauli operators. The CHSH inequality is then violated iff
\begin{eqnarray}
    \ev{\mathcal B} = 2\sqrt{m_1 + m_2} > 2,
\end{eqnarray}
where $m_{1,2}$ are the largest two eigenvalues of $M_\rho = T_\rho^{T} T_\rho$. Note that $M_\rho$ is a real symmetric matrix with squared Pauli-products in the diagonals. If we construct the matrix assuming that certain $(T_\rho)_{ij}$ Pauli products are zero, we are {\it (i)} underestimating the diagonals of $M_\rho$ and {\it (ii)} neglect some off-diagonals that cannot reduce the average of the highest two eigenvalues $m_1$ and $m_2$. Therefore estimating $\ev{\mathcal B}$ from the three simultaneously measurable Pauli products $\ev{X_1X_2}$, $\ev{Y_1Y_2}$ and $\ev{Z_1Z_2}$ is a lower bound and hence a good indicator for the violation of the inequality.

\subsection{Measuring observables of the logical Bell state using lattice surgery}
\begin{figure}
    \centering  \includegraphics[width=0.49\textwidth]{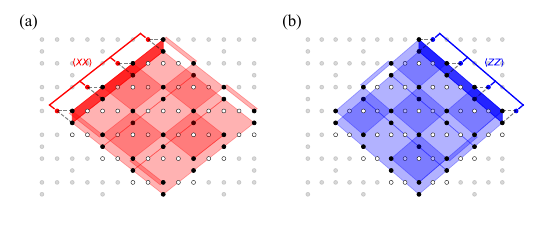}
    \caption{Measurement of the logical $XX$ and logical $ZZ$ operators in a single stabilizer round using lattice surgery on $d=4$ codes. (a) every $X$ stabilizer measurement is shown and red  qubits are the ancillas that participate in the measurement of the logical $XX$ operator shown as a non-transparent tile. (b) similarly, every $Z$ stabilizer measurement and ancilla qubits used for the readout of the logical $ZZ$ operator. For the LS measurement of the $ZZ$ observable the last ancilla in the row is not part of the lattice for the $d=4$ case. Here we applied in-place parity measurement before the CX$_\perp$ and the final readout of the logicals.}
    \label{fig:lattice_surgery}
\end{figure}

As we have seen, measuring the $XX$, $YY$ and $ZZ$ Pauli operators is sufficient to calculate the fidelity and verify the violation of Bell's inequality. Furthermore, since the Bell states only have finite expectation value with these three Paulis an approximately good state tomography can also be obtained from them. Measurement of the $Y$ observable is difficult in CSS codes defined by $X$ and $Z$ type stabilizers as it would require a logical S gate with a substantial space-time overhead. On the other hand, simultaneous $XX$ and $ZZ$ measurements also allow one to compute the $YY$ observable in a shot-by-shot basis. The simplest way to achieve this is to disentangle the logical Bell-state as in Fig.~\ref{fig:Bell_meas_prep_circuits}(c). However, this would defeat the purpose of the experiment. Namely, we want to catch the entanglement in the act, measuring correlations between logical qubits.

Lattice surgery is a well-established method to measure the $XX$ or $ZZ$ expectation values of two logical qubits. This is done by performing stabilizer measurements between two adjacent code patches, creating a connected lattice of size $2d\times d$. For BS code patches if the new stabilizer measurements connects the two patches on the edges where the logical $X$s ($Z$s) are located, the outcome of the new $X$ ($Z$) stabilizers will provide the logical observable $X_{L1}X_{L2}$ ($Z_{L1}Z_{L2}$). Usually, this method requires $d$ rounds of joint logical measurement to ensure fault tolerance against measurement errors. Combining LS with the transversal Bell measurement (as in Fig.~\ref{fig:Bell_meas_prep_circuits}(d)), we can exploit that the final measurement of the disentangled $X$ ($Z$) logical (which is on its own fault-tolerant) needs to agree with $XX$ ($ZZ$) measured in the LS scheme which lifts the aforementioned constraint. A detailed process diagram for the measurement comparisons can be found in App.~\ref{app:detector_circuits}.

The LS measurements are performed using XX or ZZ parity measurements on the edge qubits between the two codes as shown in Fig.~\ref{fig:lattice_surgery}. Similarly to the gauge measurements of the BS code, these parity measurements are performed using ancilla qubits that are located next to the edge of the two qubit patches. Each LS gauge measurement requires one conventional and one next-nearest-neighbour CX gate. We note that the new $XX$ ($ZZ$) gauge measurements do not commute with some of the original $Z$ ($X$) stabilizers. Therefore one needs to connect the ends of the $2\times d$ ($d\times 2$) stabilizer stripes forming a  $2\times 2d$ ($2d\times 2$) stabilizer in the round right after the LS measurement~\cite{gidney2023bacon}. However, during the transversal CX, we can only use product stabilizers anyway, so this peculiarity of the BS lattice surgery does not pose additional restrictions on the protocol.

Thanks to the non-planar coupling between our two codes, $XX$ and $ZZ$ lattice surgery operations can be implemented simultaneously, allowing for the extraction of $YY$. This facilitates the direct extraction of the fidelity which was only estimated in Ref.~\cite{bluvstein2023logical} from the separately performed $X$ and $Z$ measurements of the two logical qubits.
\begin{figure}[htb]
    \centering
\includegraphics[width=0.49\textwidth]{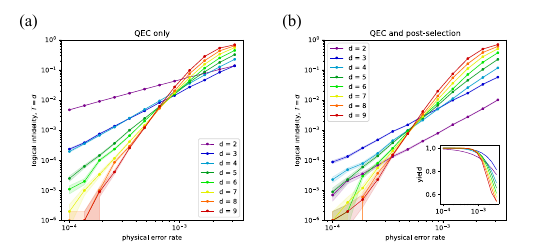}
    \caption{Infidelity $1-\mathcal{F}_\text{Bell}$ of the logical Bell state, prepared according to Fig.~\ref{fig:Bell_meas_prep_circuits}(d) with $T=d$, as a function of physical error rate. (a) Logical infidelity without post selection, and (b) same samples using post selection. The sub-threshold scaling of the even code distance curves is $d/2$ without post selection while $d/2+1$ with post selection. Inset in (b) shows the yield of the post selection.}
    \label{fig:sim_threshold_nops_wps}
\end{figure}

\section{Simulated (pseudo-)thresholds}
\label{sec:thresholds}
\begin{figure*}
    \centering
    \includegraphics[width = \textwidth]{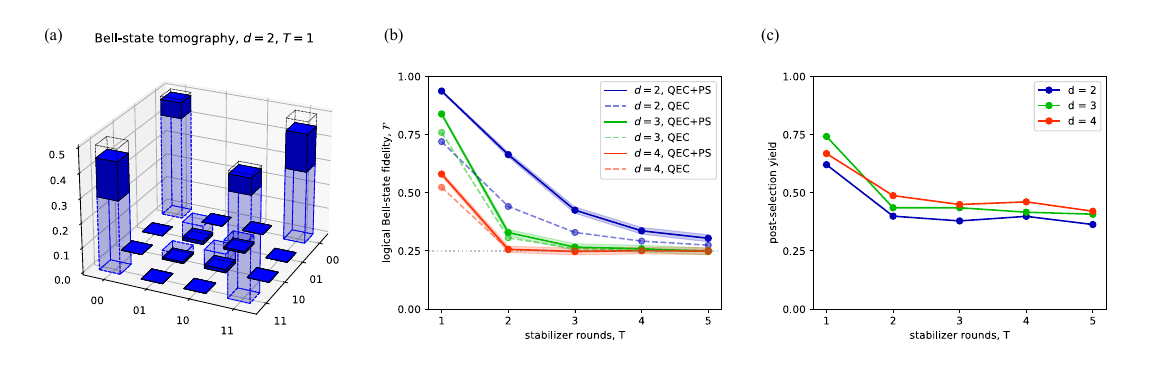}
    \caption{Bell-state preparation and measurement via transversal gates and lattice surgery. (a) measured state tomography for $d=2$ and $T=1$ without post selection (transparent bars with dashed contour), after post selection (solid bars).  (b) Bell-state fidelity after multiple stabilizer rounds with (solid) and without post-selection (dashed). Error bars correspond to $2/\sqrt{N_\text{fail}}$ uncertainty of the logical failures due to shot noise where $N_\text{fail}$ is the number of logical failures out of $50\,000$ shots. (c) post-selection yield corresponding to the lines in (b).}
    \label{fig:torino_data}
\end{figure*}

In order to verify that the LS with the transversal Bell-state measurement is indeed fault-tolerant and features the correct scaling with the code distance, we simulated the logical infidelity $1-\mathcal F_\text{Bell}$ as a function of physical error rate, under circuit level noise accounting for the decomposition of the next-nearest neighbour CX gates for the heavy-hex lattice as shown in Fig.~\ref{fig:HHXlattice}(b). In our simulations initialization, readout, single- and two-qubit gates are all assumed to be depolarizing channels with the same physical error rate $p$. For the simulations we used the same Qiskit circuits as for the real hardware experiments, converted into \texttt{stim} circuits using \texttt{qiskit-qec}. Sampling and the weighting of the decoding graph were handled by \texttt{stim} \cite{gidney2021stim} and the decoding has been performed using \texttt{pymatching} \cite{higgott2023sparse}. In order to determine the correct stabilizer comparisons for each step with nontrivial logical operations, we used the considerations discussed in Sec.~\ref{sec:detectors}.

For the simulations we used the conventional BS code without concatenations that would improve the logical error rates~\cite{gidney2023bacon}. Therefore we expect the infidelity as a function of physical error rate to have pseudo-thresholds, i.e., the crossing point of distance-$d$ and $d+2$ is higher than that of $d+2$ and $d+4$. Decoding each shot for both $XX$ and $ZZ$ logicals simultaneously, we infer $YY$ and calculate the infidelity of the logical Bell-pair. From Fig.~\ref{fig:sim_threshold_nops_wps}(a) we indeed observe pseudo-thresholds and the sub-threshold scalings seem to follow $1 - \mathcal F \sim p^{\lfloor d/2+1 \rfloor}$.

Using the post-selection scheme described in Sec.~\ref{sec:singleLQ}, we decode and post select the same samples. Indeed the scaling of the logical infidelity in Fig.~\ref{fig:sim_threshold_nops_wps}(b) improves for even distances, however the scaling of the ideal case $1 - \mathcal F \sim p^{\lceil d/2+1 \rceil}$ is not clearly reached for every code distance. Note that below the respective sub-threshold the post-selection yield remains sufficiently close to 1, and can be approximated as $1 - c_d\, p^{\lfloor d/2+1 \rfloor}$ with some distance-dependent constant $c_d$.

\section{Fidelity of logical Bell states on quantum hardware}
\label{sec:exp_fidelity}

We have tested the logical qubit protocol described in Fig.~\ref{fig:Bell_meas_prep_circuits}(d) on the latest quantum hardware, \texttt{ibm\_torino}. Circuits were executed on the cloud services using Qiskit, without any application-specific tuning of the device~\cite{kelly2016scalable,google2023suppressing}. There are a number of optimization strategies though that were employed in order to improve the logical fidelity:
\begin{itemize}
    \item interleaving the CX layers in the stabilizer measurements of the two codes, the CX schedule was chosen such that the number of CX gates and gate layers reduced by $\sim 20\%$ after transpilation;
    \item dynamical decoupling was used, i.e., a sequence of $XZX$-$XZX$ gates in every idling location in the circuit;
    \item in the final measurement round, measurement of data, ancilla, and the LS ancilla qubits are performed simultaneously (except for one LS ancilla for $d=4$).
\end{itemize}
On the other hand, we did not investigate systematically different locations for the $d=2$ and $d=3$ instances, hence better results may be achievable by choosing a location with better error rates. The chosen location for these lower code distances has been the one, where the upper corner of the respective lattice coincides with that of the $d=4$ instance (see Fig.~\ref{fig:HHXlattice}).

We performed the decoding with minimum-weight perfect-matching as described in Sec.~\ref{sec:thresholds} according to the median error rates of the device. That is a single-qubit error rate of $0.03\%$, two-qubit errors of $0.4\%$ and readout errors of $1.74\%$. Interestingly, incorporating more detailed information from the hardware, i.e., different readout error rates for vertex- and link qubits, resulted in slightly worse logical error rates, and therefore we did not pursue further optimizations in that direction. 

The result of $d=2$ and $T=1$ rounds is shown in Fig.~\ref{fig:torino_data}(a). The state tomography is obtained using the approximation $\rho_2 \approx \langle XX \rangle \sigma^x_1\sigma^x_2 + \langle YY \rangle \sigma^y_1\sigma^y_2 + \langle ZZ \rangle \sigma^z_1\sigma^z_2$. We expect that the expectation value of the Pauli products that were not measured would only slightly change the picture, since there is no indication of bias, i.e., logical states seem to converge the maximally mixed state instead of the ground state. Without post selection the logical fidelity of the $d=3$ code with $T=1$ round is $\mathcal{F}_\text{Bell} = 74.7 \pm 0.5 \%$ and the Bell's inequality is violated as $\ev{\mathcal B} = 2.05 \pm 0.02$ confirming the violation within 2$\sigma$s. Using the post selection strategy introduced above for $d=2$ and $T=1$ we obtain $\mathcal{F}_\text{Bell} = 93.7 \pm 0.04 \%$ and the expectation value $\ev{\mathcal B} = 2.63 \pm 0.01$ implies an even stronger violation of the inequality, close to the maximal value $\ev{\mathcal B}_\text{max} \approx 2.83$.

The decreasing fidelity for higher code distances in Fig.~\ref{fig:torino_data}(b) implies that the error rates of the device are above the threshold of our protocol. However, this is to be expected considering the low error threshold of the protocol (see Fig.~\ref{fig:sim_threshold_nops_wps}), which is a common property of codes that require a low-degree connectivity \cite{hetenyi2023tailoring}. In our case, the low connectivity of the device manifests itself not in the QEC code but in the large overhead of the next-nearest-neighbour two-qubit gates ($N_\text{CX} \sim 32d^2T$). 

Logical error rate after a single stabilizer round paints an overly optimistic picture of the logical error rate. This is due to the fact, that the first stabilizer round projects a product state into the code space rather than probing the logical qubit that is already in the code space. In order to gain a better insight into the decoherence of the logical qubit during stabilizer measurements, we present the decay of logical fidelity for code distances $d\in\{2,3,4\}$ over multiple rounds in Fig.~\ref{fig:torino_data}(b)-(c). Together with the post-selected data, we also show the logical Bell-state fidelity with QEC only, and provide the post-selection yield (for an additional data set in App.~\ref{app:additional_data}).

\section{Conclusion}

We have realized two different topological codes the 3CX and the Bacon-Shor code in the heavy-hex lattice. Using transversal CX gate and simultaneous $XX$ and $ZZ$ lattice surgery we entangled the logical qubits of the two codes for the first time in the superconducting qubit platform. Moreover, we have have verified the entanglement between the two logical qubits over multiple code distances and stabilizer measurement rounds. Our results show violation of Bell's inequality implying that non-classical correlations can be restored using quantum error correction.

Our results also allow the 3CX code to be compared to the heavy-hex code~\cite{chamberland2020topological}, which has also been implemented on IBM Quantum hardware~\cite{chen2022d2,sundaresan2023d3}. Comparable results show largely similar performance. The sparse implementation of the 3CX code therefore does not suffer significantly in comparison to a code tailored to the connectivity.

Finally, it is worth noting that non-planar coupling is also possible in solid state approaches, as would be required for high-rate LDPC codes~\cite{bravyi2024highthreshold}. Nevertheless, we hope that our techniques will inspire even the most sophisticated coupling schemes to make the most out of what they have.

\section*{Data availability}

The python scripts used to generate the \texttt{Qiskit} and \texttt{stim} circuits as well as those for the decoding and post selection are available at \cite{bellgithub}.

\section*{Acknowledgements}

The authors would like to thank Ben Brown and Ian Hesner for helpful discussions. We are grateful for the donation of IBM Quantum resources to complete this project. BH acknowledges support from the NCCR SPIN, a National Centre of Competence in Research, funded by the Swiss National Science Foundation (grant number 51NF40-180604).

\appendix

\section{Threshold plots of the 3CX and Bacon-Shor codes with heavy-hex connectivity}
\label{app:more_thresholds}
\begin{figure}
    \centering\includegraphics[width=0.49\textwidth]{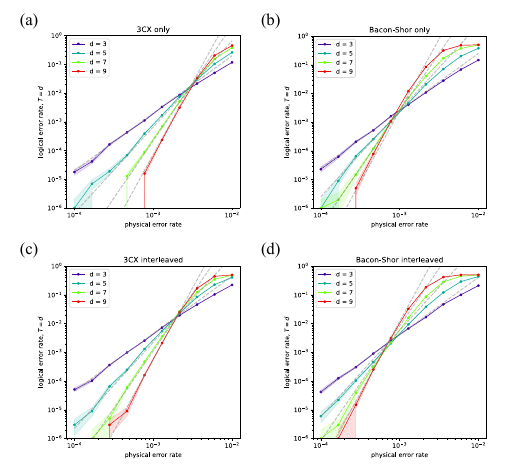}
    \caption{Logical error rate as a function of physical error rate for the $\ket{0}_L$ state of the 3CX and BS codes. (a)-(b) individual threshold plots of the 3CX and BS codes, respectively. (c)-(d) threshold plots of the 3CX and BS logical qubits, respectively, when the two codes are executed simultaneously.  Simulations were performed under circuit-level noise assuming heavy-hex connectivity. Dashed lines correspond to $P_L = P_{L,0} (p/p_\text{th})^{(d+1)/2}$ for the plotted code distances.}
    \label{fig:3CXBSsim_threshold_nops}
\end{figure}

Under circuit-level noise, accounting for the CX overhead of the next-nearest-neighbour CX gates we obtain approximate threshold rates of $p_\text{th} = 0.3\%$ and $\tilde p_{(3,9)} \approx 0.1\%$ for the 3CX and BS codes, respectively (see Fig.~\ref{fig:3CXBSsim_threshold_nops}). Here $\tilde p_{(3,9)}$ indicates an averaged pseudo-threshold. Similarly, when the two codes are executed simultaneously with interleaved CX gate layers we find $p_\text{th} = 0.2\%$ and $\tilde p_{(3,9)} \approx 0.06\%$ for the 3CX and BS codes, respectively. Importantly, the sub-threshold scaling is maintained also for the interleaved case, indicating that the next-nearest neighbour CX gates do not reduce the code distance of either of the codes.

\section{Device benchmarks}
\label{app:RBdata}
\begin{table}[htp]
\begin{center}
\begin{tabular}{|c|c|c|}
\hline
 \rule{0pt}{10pt}   & vertex qubits & link qubits \\[2pt]
\hline
 \rule{0pt}{10pt} CZ error [\%] & \multicolumn{2}{c|}{0.54 (0.38)} \\[2pt]
\hline
 \rule{0pt}{10pt} SX error [\%] & 0.033 (0.026) & 0.047 (0.037)\\[2pt]
\hline
 \rule{0pt}{10pt} Readout error [\%] & 3.4 (3.1) & 1.1 (1.0) \\[2pt]
\hline
 \rule{0pt}{10pt} $\tau_\text{RO}/T_1$ [\%] & 1.3 (0.87) & 0.93 (0.74) \\[2pt]
\hline
 \rule{0pt}{10pt} $\tau_\text{RO}/T_2$ [\%] & 1.7 (1.2) & 1.3 (1.2) \\[2pt]
\hline
\end{tabular}
\end{center}
\caption{Average (median) values for the qubits and qubit pairs participating in the $d=4$ codes. SX refers to the single-qubit square-root-X gate and $\tau_\text{RO}$ is the integration time of the readout.}
\label{tab:RB_data}
\end{table}

The device benchmarks summarized in Tab.~\ref{tab:RB_data} correspond to the calibration data of \texttt{ibm\_torino} on the 11th February, 2024, when the data has been taken. It is worth noting that the readout errors are considerably worse on the vertex qubits. Furthermore, errors on idling qubits are not accurately described by $T_1$ and $T_2$, due to the `XZX-XZX' dynamical decoupling sequence applied on the data qubits while idling.

\section{Additional data}
\label{app:additional_data}
\begin{figure*}
    \centering
    \includegraphics[width = \textwidth]{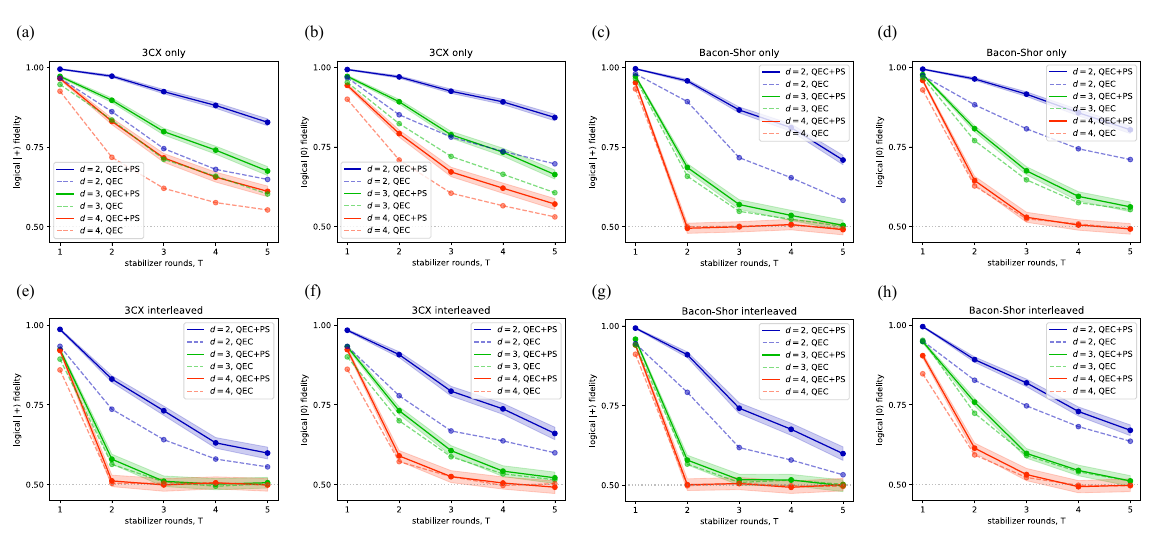}
    \caption{Preparation and measurement of the $\ket{+}$ and $\ket{0}$ states in the 3CX and Bacon-Shor codes. (a)-(d) measuring stabilizer outcomes of the 3CX (BS) code only with next-nearest-neighbour CX gates while the link (vertex) qubits are initialized in the beginning. (e)-(h) simultaneous execution of the two codes with preparation of the $\ket{0}_\text{3CX}\ket{+}_\text{BS}$ and $\ket{+}_\text{3CX}\ket{0}_\text{BS}$ states.
    Error bars correspond to $2\sqrt{N_\text{fail}}$ uncertainty of the logical failures due to shot noise where $N_\text{fail}$ is the number of logical failures out of $10\,000$ shots.}
    \label{fig:torino_data_3CX_BS}
\end{figure*}

\begin{figure}
    \centering
    \includegraphics[width = 0.49\textwidth]{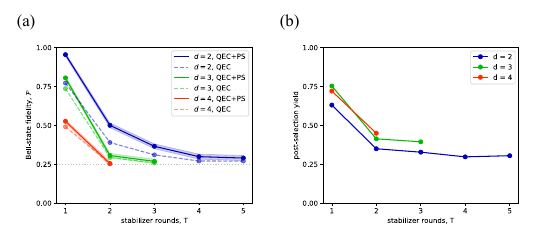}
    \caption{Second data set for the logical Bell-state experiment. (a) Bell-state fidelity after multiple syndrome extraction rounds with (solid) and without post-selection (dashed). Error bars correspond to $2\sqrt{N_\text{fail}}$ uncertainty of the logical failures due to shot noise where $N_\text{fail}$ is the number of logical failures out of $50\,000$ shots. (b) post-selection yield. Missing data points are results of buffer overflows during result collection, due to the high number of measurement.}
    \label{fig:torino_data_old}
\end{figure}

Here we present additional data taken on the same device (\texttt{ibm\_torino}) for isolated and interleaved 3CX and BS codes as well as for logical qubits prepared in the $\ket 0$ and $\ket +$ states as shown in Fig.~\ref{fig:torino_data_3CX_BS}. The location of the smaller-distance codes have been chosen to match at the upper corner with the $d=4$ variant shown in Fig.~\ref{fig:HHXlattice}.

First, we analyzed performance of the 3CX code, without link qubits, i.e., initializing the link qubits in the state $\ket{0}$ in the beginning. Performing a quantum memory experiment on the logical $\ket{0}_L$ and  $\ket{+}_L$ state produced encouraging results with non-trivial logical fidelity up to $d=4$ and $T=5$ (see Fig.~\ref{fig:torino_data_3CX_BS}(a)-(b)).

Next, we studied the Bacon-Shor code on the link qubits separately (vertex qubits initialized in $\ket 0$). While the results for lower code distances are good enough to build on them, there is a substantial difference between the $\ket{0}_L$ and $\ket{+}_L$ states, especially for higher code distances (see Fig.~\ref{fig:torino_data_3CX_BS}(c)-(d)).

Furthermore, we have repeated the experiments above, using the interleaved CX layers and executing the stabilizer measurements of the two codes simultaneously. Respecting the order of the CX gates in the 3CX code, we found that the lowest number of entangling gates and gate layers is obtained for the following schedule: 3CX round-0, BS Z-rounds 0 and 1, 3CX rounds 1 and 2, BS X-rounds 0 and 1, and 3CX round-3. This reduced the gate count to $\approx 80\%$ compared to the subsequent evaluation of the 3CX and BS next-nearest-neighbour CX layers. Results for both $\ket{0}_L$ and $\ket{+}_L$ initial states have been obtained preparing the two logical qubits either in the $\ket{0}_\text{3CX}\ket{+}_\text{BS}$ states or in the $\ket{+}_\text{3CX}\ket{0}_\text{BS}$ states (see Fig.~\ref{fig:torino_data_3CX_BS}(e)-(h)). For the interleaved case, the post-selected results are only plotted for reference. Our post-selection strategy --that is based on the ambiguity of matching to different boundaries-- does not increase the code distance in this case, since there are internal edges in the decoding graph (i.e., edges that are not connected to any boundary node) that also imply logical flip(s).

Finally, we show an additional data set taken at an earlier time (January 13th, 2024) on Fig.~\ref{fig:torino_data_old}. This data set shows similar performance to that of Fig.~\ref{fig:torino_data}. For this dataset the the location of the $d=2,3$ instances are such that they match on their left corner with the $d=4$ case.

\section{Logical CX and hyperedges in the decoding graph}
\label{app:hyperedges}

While, considering the codespace, it is straightforward to show that the CX gate can be applied transversally for two BS qubits, the detectors, i.e., comparison of stabilizer measurements, need to be adjusted in the round that incorporates the transversal operation~\cite{bluvstein2023logical}. Here we use the detecting region picture introduced by Ref.~\cite{mcewen2023relaxing}, which can be considered as a pictorial representation of how the stabilizer group evolves upon the application of physical gates. 

\begin{figure}
     \centering
    \includegraphics[width=0.49\textwidth]{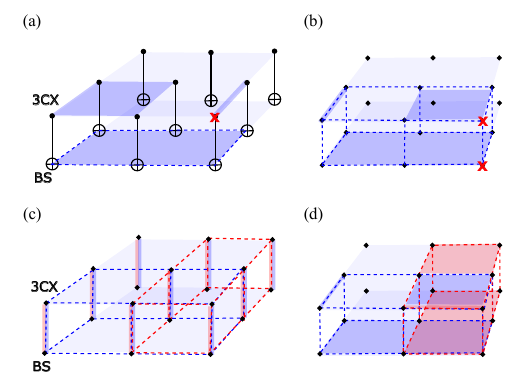}
    \caption{
    (a)-(b) transversal CX operation controlled on the BS LQ and targeted on the 3CX LQ. Z-type error detecting regions are defined as the product of $Z$ stabilizers encircled by the dashed lines. An $X$ error propagating between lattices are shown with red cross. (c)-(d) transversal Bell-state preparation and an $X$ ($Z$)-type inter-qubit stabilizer that is an error detection event in the first round shown as red (blue) dashed line.}
    \label{fig:det_regs}
\end{figure}

Let us consider a $Z$ stabilizer outcome of the BS code and an appropriate set of stabilizer outcomes of the 3CX code at a given time $t$ as in Fig.~\ref{fig:det_regs}(a). The qubitwise CX gates will extend the dashed blue detecting region from the BS side to the 3CX side of the two LQ system. Consequently, the BS stabilizer at time $t$ needs to be compared not only to that of $t+1$ but also to the BS-like subset of stabilizer measurements at round $t+1$ of the 3CX code. This process is shown in Fig.~\ref{fig:det_regs}(a)-(b) along with the propagation of $X$ errors which helps verifying that the expanded BS region does not signal an error that is transplanted from the 3CX code during the transversal CX. This picture is equivalent to multiplying certain measurements into the detection events from the previous round as suggested by Ref.~\cite{bluvstein2023logical}.

Note that $Z$ ($X$) stabilizers of the 3CX (BS) code are unaffected by the CX gates, validating the respective stabilizer comparisons. Therefore the original $Z$ ($X$) detecting regions will overlap with the extended ones, discussed above. The overlapping structure of detecting regions gives rise to hyperedges in the decoding graph. E.g., 3CX (BS) qubits that are not at the edge of the lattice will be covered by four $Z$ ($X$)-type detecting regions, implying that an error in the appropriate location in the circuit results in four detection events, i.e., a hyperedge connecting four nodes in the decoding graph. Depending on the number of rounds, the code distance, and the error model these hyperedges can be decomposed by \texttt{stim}. E.g., hyperedges can be decomposed for $T=2$ for any code distance if the logical CX is applied after the first round of measurements, or for an arbitrary setting for $d=3$ if two-qubit errors are allowed (hook errors create effectively non-local syndromes, as opposed to $d\geq4$ where these are only 2-local). The smallest example where a single readout error triggers a hyperedge is $T=3$, logical CX at $T=1$ for any $d\geq 4$.

\section{Transversal Bell-state preparation and measurement}
\label{app:detectiong_regions}

Here we consider the detecting regions associated with the transversal preparation and measurement of a logical Bell pair and show that their decoding problem is compatible with a conventional matching decoder. 

As discussed in App.~\ref{app:hyperedges}, detecting regions overlap after the CX$_\perp$ gates. However, detecting regions, similar to stabilizer operators, are generators of a group with the product operation, implying that the product of overlapping region is also a valid detecting region. Therefore, an equivalent choice of detecting region in Fig.~\ref{fig:det_regs}(a)-(b) would be a contracting $Z$ ($X$) region that terminates on the BS (3CX) side. 

This raises the question, what happens if the CXs are applied before the first or after the last round of regular stabilizer measurements? In the former case, the CXs couple unentangled qubits rather than LQs. This case is interesting nonetheless, since it can be used to prepare a logical Bell state. If every 3CX (BS) qubits have been initialized on the $X$ ($Z$) basis, the qubitwise CXs prepare $d^2$ Bell states each of which are eigenstates of the $XX$ and $ZZ$ stabilizers acting on the respective pairs of physical qubits (see Fig.~\ref{fig:det_regs}(c)). Consequently, the product of compatible 3CX and BS stabilizers are deterministic in the first round such that two (one) detecting regions of each type cover every bulk (edge) qubit (see Fig.~\ref{fig:det_regs}(d)).

Importantly, none of the stabilizers of the individual codes commute with all of the Bell-pair stabilizers of the initial stabilizer group. Only the brick-shaped regions are valid detecting regions meaning that at most two regions cover each qubit leading to regular edges in the decoding graph. After the first round, the regular stabilizer outcome comparisons can be performed for both of the codes. Similarly, applying the qubitwise CXs before the final $X$ and $Z$ measurements of the 3CX and BS LQs, implements a logical measurement on the Bell basis. Note that any other final measurement (e.g., $X_{L1}$ and $X_{L2}$) will continue to have the hyperedge problem.

\section{Detectors during the Bell-state preparation and lattice surgery}
\label{app:detector_circuits}
\begin{figure}
    \centering
    \includegraphics[width=0.49\textwidth]{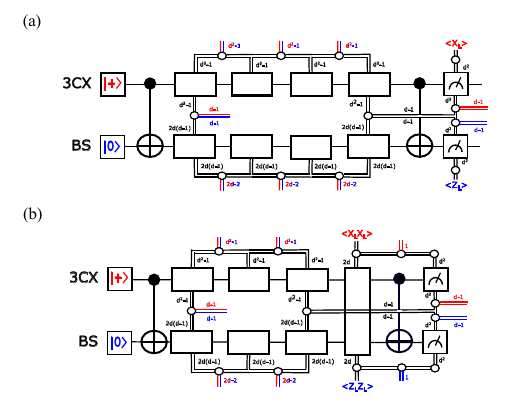}
    \caption{Process diagram for the stabilizer measurement comparisons. Red (blue) lines correspond to detectors of the $X$ ($Z$) type. (a) detectors for logical Bell state preparation, four rounds of stabilizer measurements, and final readout on the Bell basis. (b) detectors for one stabilizer round followed by a lattice-surgery measurement of the correlated Bell-state observables ($XX$ and $ZZ$) and final measurement on the Bell basis.}
    \label{fig:Bell_LS_circuits}
\end{figure}

As we described in App.~\ref{app:detectiong_regions}, the compatible products of 3CX and BS stabilizer measurements are deterministic in the first round if we start by entangling physical qubit pairs. Since we prepare every Bell pair in the same state, only the stabilizer outcomes in the first stabilizer measurements need to be compared to detect errors in the initialization. This is represented on the leftmost colored lines in Fig.~\ref{fig:Bell_LS_circuits}(a). In total there are $d-1$ such detectors from each type which is a reminiscent from the number of stabilizers in the BS code. 

In subsequent stabilizer rounds, stabilizer comparisons can continue according to the respective codes. As opposed to a quantum memory experiment on a single logical qubit, here we need to keep of track both types of stabilizers to identify both bit- and phase-flip errors in the logical Bell state. In the last round of stabilizer measurements, however, we can only use the $2(d-1)$ product stabilizers to compare them with appropriate products of the final Bell-pair measurements. 

The former scheme only allows us to measure the logical observables after they are disentangled. However, performing a lattice surgery the $\ev{X_{L1}X_{L2}}$ and $\ev{Z_{L1}Z_{L2}}$ logical observables can be measured without measuring the logical qubits separately (see the corresponding diagram in Fig.~\ref{fig:Bell_LS_circuits}(b)). This can be considered as a straightforward extension of the previous case, in a sense that we obtain the same information before (e.g., $\ev{X_{L1}X_{L2}}$) and after (e.g., $\ev{X_{L1}}$) disentangling the logical qubits. Importantly, we can choose the LS measurement outcomes (that are on their own not fault-tolerant after a single round of measurements) and compare them with the final measurement of the disentangled observables that is fault-tolerant. This latter comparison will be an additional detector that tells us if the LS measurement needs to be flipped or not.

\bibliography{references}

\end{document}